\documentclass{appolb}
\usepackage{epsfig}
\usepackage{amsmath}
\usepackage{amssymb}

\begin{document}

\title{\bf On statistical mechanics of a single particle in
high-dimensional random landscapes \footnote{The text is based on
presentations at meetings "Random Matrix Theory: from fundamental
physics to applications", May 2-6, 2007, Krakow, Poland and
"Condensed Matter Physics meets High-Energy Physics", July 13-18,
2007, St. Petersburg, Russia.}}

\author{Yan V. Fyodorov
\address{School of Mathematical Sciences,
University of Nottingham , Nottingham NG72RD, United Kingdom}}

\maketitle

\begin{abstract}
We discuss recent results of the replica approach to statistical
mechanics of a single classical particle placed in a random $N
(\gg 1)$-dimensional Gaussian landscape. The particular attention
is paid to the case of landscapes with logarithmically growing
correlations and to its recent generalisations. Those landscapes
give rise to a rich multifractal spatial structure of the
associated Boltzmann-Gibbs measure. We also briefly mention
related results on counting stationary points of random Gaussian
surfaces, as well as ongoing research on statistical mechanics in
a random landscape constructed locally by adding many squared
Gaussian-distributed terms.

\end{abstract}


\vskip 0.5cm

One of the simplest models with quenched disorder - a single
classical particle subject to a superposition of random Gaussian
potential $V({\bf x})$ and a non-random confining potential
$V_{con}({\bf x})$, with ${\bf x}\in \mathbb{R}^N$ - turns out to
be a surprisingly rich system, characterised by a non-trivial
dynamical behaviour as well as interesting thermodynamics.
Denoting the total potential energy as ${\cal H}({\bf
x})=V_{con}({\bf x})+V({\bf x})$, the statistical mechanics of the
model is controlled by the free energy:
\begin{equation}\label{freeendef}
F_N=-\beta^{-1}\,\langle \ln{Z(\beta)}\rangle_V ,\, \quad
Z(\beta)=\int_{|{\bf x}|\le L} \exp{-\beta {\cal H}({\bf x})}\, d
{\bf x}\,
\end{equation}
as a function of the inverse temperature $\beta={1}/{T}$, and the
sample size $L$, with brackets standing for the averaging over the
Gaussian potential distribution. The covariance function of the
random part is usually chosen in the form ensuring stationarity
and well-defined large-$N$ limit:
\begin{equation}\label{1}
\left\langle V\left({\bf x}_1\right) \, V\left({\bf,
x}_2\right)\right\rangle_V=N\,f_V\left(\frac{1}{2N}({\bf x}_1-{\bf
x}_2)^2\right)\,.
\end{equation}

Important information about the structure of the Gibbs-Boltzmann
equilibrium measure $p_{\beta}({\bf
x})=\frac{1}{Z(\beta)}\exp{-\beta {\cal H}({\bf x})}\,$ can be
extracted from the knowledge of moments
\begin{equation}\label{BGmom}
\quad m_q=\int_{|{\bf x}|\le L} p^q_{\beta}({\bf x})\, d {\bf
x}=\frac{Z(\beta q)}{\left[Z(\beta)\right]^q}\,.
\end{equation}
In the thermodynamic limit of the sample volume $V_L \to \infty$
one expects typically
\begin{equation}\label{BGmom1}
\quad m_q\sim V_L^{-\tau_q}\,,
\end{equation}
where the set of exponents $\tau_q$ reflects the spatial
organization of the Gibbs-Boltzmann weights. For example, if the
weights are of the same order of magnitude across the sample
volume, the normalization condition implies locally
$p_{\beta}({\bf x})\sim V_L^{-1}$ and a simple power counting
predicts the exponents $\tau_q=q-1$. In such a situation it is
conventional to speak about a {\it delocalised} measure. The
opposite case of a fully {\it localised} measure describes the
situation when essential Gibbs-Boltzmann weights concentrate in
the thermodynamic limit in a domain with the finite total volume
$V_{\xi}\ll V_L\to \infty$, and are vanishingly small outside that
domain. This situation is obviously characterized by trivial
exponents $\tau_{q>0}=0$ and $\tau_{q<0}=\infty$. Finally, in many
interesting situations the exponents $\tau_{q}$ may depend on $q$
nonlinearly, and in this case one commonly refers to the {\it
multifractality} of the measure. The Eqs.(\ref{BGmom}) and
(\ref{BGmom1})
 imply the following expression for the characteristic exponents $\tau_q$
 in the general case
\begin{equation}\label{BGmom2}
\quad \tau_q=|q|\beta{\cal F}(|q|\beta)-q\beta{\cal F}(\beta)
\end{equation}
relating them to the appropriately normalized free energy of the
system:
\begin{equation}\label{BGmom3}
{\cal F}(\beta)=-\lim_{V_L\to \infty}\frac{\ln{Z(\beta)}}{\beta
\ln{V_L}}\,.
\end{equation}

The investigation of multifractal measures of diverse origin is a
very active field of research in various branches of physics for
more than two decades. In recent years important insights were
obtained for disorder-generated multifractality, see \cite{ME} for
a comprehensive discussion in the context of the Anderson
localization transition, and also \cite{MG} for an example related
to statistical mechanics with disorder. The multifractality of
random Gibbs-Boltzmann measures in a context related to ours
appeared in the insightful paper \cite{2d}. A popular way of
characterizing multifractality invokes the so-called singularity
spectrum function $f(\alpha)$. The latter function is used to
characterize the number $dN(\alpha)=V_L^{f(\alpha)}d\alpha$ of
sites in the sample where the local Gibbs-Boltzmann measure scales
as $p_{\beta}({\bf r})\sim V_L^{-\alpha}$ in the thermodynamic
limit. The definition allows to extract the typical characteristic
exponents $\tau_q$ as, see e.g. \cite{ME}
\begin{equation}\label{BGmom4}
\tau_q=-\lim_{V_L\to \infty}\frac{\ln{\int_{f(\alpha)\ge
0}e^{-\ln{V_L}[\alpha q-f(\alpha)]}\,d\alpha}}{\ln{V_L}}\,.
\end{equation}
Performing the $\alpha-$integration by the Laplace method we
obtain that the positive values of the multifractality spectrum
$f(\alpha)$ are related by the Legendre transform to the set of
exponents $\tau_q$:
\begin{equation}\label{sinspec}
\tau_q= \alpha_* q-f(\alpha_*), \quad q=f'(\alpha_*),
\end{equation}
Thus, the knowledge of the free energy ${\cal F}(\beta)$ in
(\ref{BGmom3}) allows one to characterize the positive part of the
multifractality spectrum of the Boltzmann-Gibbs measure.

Early works by Mezard and Parisi \cite{MP}, and Engel \cite{Engel}
used the replica trick to calculate the free energy
Eq.(\ref{freeendef}) of an infinite system, $L=\infty$, confined
by the simplest parabolic potential $V_{con}({\bf
x})=\frac{1}{2}\mu{\bf x}^2,\,$$\mu>0$. Employing the so-called
Gaussian Variational Ansatz (GVA) the authors revealed the
existence of a low-temperature phase with broken replica symmetry,
hence broken ergodicity. They were followed by Franz and Mezard
\cite{FM} and Cugliandolo and Le Doussal \cite{longtime} papers on
the corresponding dynamics revealing long-time relaxation, aging,
and other effects typical for glassy type of behaviour at low
enough temperatures. The nature of the low-temperature phase was
found to be very essentially dependent on the type of correlations
in the random potential, specified via the covariance function
described in Eq.(\ref{1}).
 Namely, if the covariance $f_V(u)$ decayed
to zero at large arguments $u$, the description of the low
temperature phase was found to require only the so-called one-step
replica symmetry breaking (1RSB) Parisi pattern. This effect
correctly captures the statistics of the low-lying minima of
associated Gaussian energy landscapes\cite{BM}.

In contrast, for the case of long-ranged correlated potentials
with $f_V(u)$ growing with $x$ as a power-law\footnote{To be more
precise, at large separations we require the structure function
$\left\langle \left[V\left({\bf x}_1\right)- V\left({\bf
x}_2\right)\right]^2\right\rangle_V\propto \left({\bf x}_1-{\bf
x}_2\right)^{2\gamma}, \, 0<\gamma<1$. However one can easily
satisfy oneself that in the present model under consideration the
difference between the structure function and the covariance is
immaterial for the free-energy calculations. This will be no
longer the case for the model discussed in the end of this
article.} the full infinite-hierarchy Parisi scheme of replica
symmetry breaking (FRSB) had to be used instead.

Based on formal analogies with the Hartree-Fock method Mezard and
Parisi\cite{MP} argued that GVA-based calculations should become
exact in the limit of infinite spatial dimension $N$. In a recent
paper \cite{FS} the replicated problem was reconsidered in much
detail by an alternative method which directly exposed the degrees
of freedom relevant in the limit $N\to \infty$, and in this way
allowed to employ the Laplace (a.k.a. saddle-point) evaluation of
the integrals. The results obtained in \cite{FS} by this method
for the parabolic confinement case fully reproduced those obtained
by GVA in \cite{MP,Engel}.

The method of \cite{FS} also works for statistical mechanics of a
single particle inside any spherical sample $|{\bf x}|<L$ of a
given radius $L$ which makes it particularly suitable for
studying, e.g., multifractality of the associated Gibbs-Boltzmann
measure. To this end it is easy to understand that the radius $L$
must be scaled with the dimension as $L=R\sqrt{N}$ to ensure
nontrivial results when $N\to \infty$. The effective size
$R<\infty$ (which is actually half of the length of an edge of the
cube inscribed in this sphere) can be used as an additional
control parameter of the model. In particular, the chosen scaling
$L=R\sqrt{N}$ ensures that the sample volume
$V_L=\pi^{N/2}\frac{L^N}{\Gamma(N/2+1)}$ retains in the limit
$N\to \infty$ the natural scaling with size $R$ and dimension $N$.
Namely, for $R\gg 1$ we have $\ln{V_L}=N\ln{R}+$ smaller terms,
which is very essential for the analysis of multifractality.

 In what follows we thus concentrate on the case of no confinement
potential $V_{con}({\bf x})=0$, and choose $R$ to have any fixed
value (eventually we will be interested in a kind of thermodynamic
limit $R\to \infty$). One of the observations made in \cite{FS} is
the existence of a simple mathematical criterion which formally
differentiates between the short-range correlated potentials and
their long-ranged counterparts. Namely, assume the covariance
function $f_V(u)$ in (\ref{1}) to satisfy technical conditions
$f'_V(u)<0,\, f''_V(u)>0$ and $f'''_V(u)<0$ for all $u\ge 0$, and
also $f_V'(u)\to 0$ when $u\to \infty$. The criterion is based on
considering a combination $A(u)$ expressed in terms of $f(u)$
as\footnote{ This eventually coincides, up to the overall sign,
with the standard definition of the so-called Schwarzian
derivative $\{f'(u),u\}$.}
\begin{equation}\label{2}
A(u)=\frac{\frac{3}{2}\left[f_V'''(u)\right]^2-f_V''(u)f_V''''(u)}{[f_V''(u)]^2}\,,
\end{equation}
where dashes indicate the order of derivatives taken. Then any
potential satisfying $A(u)>0,\,\forall u\ge 0$ (this family
includes, e.g., the potentials with
$f_V(u)=\exp{[-(a+bu)^{\alpha}]}$, such that $a>0,b>0$ and
$0<\alpha\le 1$) turned out to have the low-temperature phase
which is necessarily of 1RSB type. The standard replica stability
analysis of this 1RSB low-temperature phase revealed that the
stability is controlled by two eigenmodes, denoted in \cite{FS} as
$\Lambda_0^*$ and $\Lambda_K^*$ ( see equations (B.29) and (B.30)
of the Appendix B of that paper). If both are positive, all other
eigenvalues of the stability matrix are positive and the 1RSB
solution corresponds to an extremum of the free energy functional
stable with respect to small variations. And those two eigenvalues
were indeed found to be strictly positive as long as $A(u)>0$.

The situation was found to be very different for the potentials
with $A(u)<0,\,\forall u\ge 0$ (this family includes, most
notably, the powerlaw-correlated potentials with the covariance of
the form $f_V(u)=f(0)-g^2(u+a)^{\gamma},\, f(0)>g^2\,a^\gamma,\,
0<\gamma<1$). The low temperature phase is now of FRSB type, and
it is only "marginally stable". Indeed, the stability matrix for
this type of the replica symmetry breaking can be shown to contain
always a family of zero "replicon" modes, see e.g. \cite{CDT} for
a calculation in the framework of GVA.

Finally, the above criterion naturally singles out the random
potentials satisfying $A(u)=0, \,\forall u\ge 0$ as a boundary
case between the two regimes. Denoting $\tilde{f}(x)=f_V''(u)$ and
noticing that $A(u)=0$ implies $\tilde{f}'/\tilde{f}^{3/2}=const$,
we find the function $f_V(u)$ to be equal to
$f_V(u)=C_0^{-2}\ln{(C_0u+C_1)}+C_2u+C_3$, where $C_i$ are
arbitrary constants. The condition $f_V'(u)\to 0$ when $u\to
\infty$ then selects the case of logarithmic correlations as the
only possible, which we write as
\begin{equation}\label{2c}
f_V(u)=f_0-g^2\ln{(u+a^2)}\,.
\end{equation}

This latter choice turns out to be in many respects the most
interesting situation. Indeed, in \cite{FS} it was shown that it
leads to a phase diagram which combines features typical for the
short-ranged behaviour with others characteristic of the
long-ranged disorder. As a particular interesting feature we would
like to mention that although the low-temperature phase can be
thought of as described by a special case of 1RSB breaking scheme,
the relevant eigenvalues $\Lambda_0^*$ and $\Lambda_K^*$ of the
stability matrix identically vanish everywhere in the
low-temperature phase\footnote{ This fact, though not explicitly
mentioned in \cite{FS}, immediately follows from definitions
(B.29) and (B.30) after substituting for $q_1-q_0=Q$ and
$q_d-q_1=y$ the expressions (74) and (79) of that paper.},
rendering 1RSB phase in this special case marginally stable.

The qualitative difference between the three cases - short-ranged,
long-ranged, and logarithmic, is most clearly seen in the
thermodynamic limit of large sample size $R\to \infty$. One finds
that for a typical short-ranged potential the domain of existence
of 1RSB phase vanishes as long as $R\to \infty$. For example, the
transition (de-Almeida-Thouless\cite{AT}, AT) temperature
signalling of instability of the replica-symmetric solution
typically behaves as $T_{AT}(R)\approx R^2\sqrt{f_V''(R^2)}$ and
rapidly tends to zero for decaying correlations. For any fixed
temperature $T>0$ in the limit $R\to \infty$ the system is
effectively in the high-temperature replica symmetric phase, and
the free energy behaves asymptotically like $F(T)\approx
-T\,N\ln{R}$. Actually this result can be seen as a purely
entropic contribution, and in particularly implies via
Eqs.(\ref{BGmom2},\ref{BGmom3}) the trivial scaling of the
exponents $\tau_q=q-1$, corresponding to the totally delocalized
Boltzmann-Gibbs measure.

In contrast, for a power-law growth of correlations one finds that
the low-temperature glassy phase occupies bigger and bigger
portion of the phase diagram with growing radius $R$. Indeed, the
transition temperature can be shown to grow with $R$ as
$T_{AT}(R)\sim R^{\gamma}$, and increasing the system size $R$ at
any fixed temperature $T>0$ results in the free energy given
asymptotically by the temperature-independent value $F(T)|_{R\to
\infty}\sim -NR^{\gamma}$. This expression actually coincides with
the typical minimum of the energy function for our system. The
corresponding exponents $\tau_{q>0}=0$. In a sense the system of
this type is always "frozen" in the thermodynamic limit, and
indeed the Boltzmann-Gibbs measure is localized on a few deep
minima.

Only for the logarithmic case Eq.(\ref{2}) the transition
temperature tends in the thermodynamic limit to a finite value
$T_{AT}(R\to \infty)=g$ , and the free energy asymptotics depends
non-trivially on the temperature:
\begin{equation}\label{REM}
F(T)|_{R\to \infty}\approx -N\,\ln{R}\,\left\{\begin{array}{l} T(1+g^2/T^2),\quad T>g\\
2g,\qquad\qquad\qquad T<g
\end{array}\right.
\end{equation}
This is natural to interpret as a freezing transition, precisely
of the same sort as appeared in the celebrated Random Energy Model
(REM) by Derrida\cite{Derrida}. The same expression for the free
energy appeared actually in studies of a zero-energy wavefunction
for Dirac particles in dimension $N=2$ and random magnetic field
\cite{2d}, after a mapping to a problem of statistical mechanics.
The Boltzmann-Gibbs measure in this particular case is
characterized via a set of non-trivial multifractality exponents
$\tau_q$, see \cite{2d} and also \cite{CL}. The ensuing
multifractality spectrum $f(\alpha)$ is simple parabolic for all
temperatures, and shows interesting "freezing" behaviour for
$\alpha\to 0$ at $T=g$, i.e. at the point of ergodicity breaking.

We thus see that our results have counterparts in the
finite-dimensional systems.  Actually, understanding the generic
statistical-mechanical behaviour of disordered systems for finite
$N$ remains very challenging problem. To this end, rather detailed
attempt of investigating our model for finite dimensions
$N<\infty$ in the thermodynamic limit $L\to \infty$ was undertaken
in a very insightful paper by Carpentier and Le Doussal \cite{CL}.
That paper also can be warmly recommended for describing the
present model in a broad physical context and elucidating its
relevance for quite a few other interesting and important physical
systems, as e.g. directed polymers on trees \cite{DS}. The work
was based on employing a kind of real-space renormalisation group
(RG) treatment augmented with numerical simulations. The authors
concluded that for finite spatial dimensions neither models with
short-ranged, nor with long-ranged correlations can display a true
phase transition at finite temperatures $T>0$.  And only if
correlations grow {\it logarithmically} with the distance, for
such marginal situation the true REM-like freezing transition
indeed happens at some finite $T>0$ at any dimension $N\ge 1$.
Fortunately, the logarithmic growth is not at all an academic
oddity, but rather emerges in quite a few systems of actual
physical interest, see \cite{CL} for a detailed discussion and
further references.

We thus see that the picture following from the results of
\cite{FS} for the thermodynamic limit (understood as $R\to
\infty$) of the model in infinite dimension is in overall
qualitative agreement with the $N<\infty$ renormalization group
studies of the same model in the limit $L\to \infty$. Another fact
which is perhaps worth mentioning is that a recent work
\cite{Moore} claimed that 1RSB low-temperature phase fails to
survive in finite spatial dimensions, the fact being related to
absence of marginally stable modes in the fluctuation spectrum. If
one assumes that the validity of that claim extends beyond the
particular model considered in \cite{Moore}, then in our case 1RSB
phase in finite dimensions has no chance of survival for any
short-range potentials, but in the logarithmic case it could
survive due to the mentioned marginal stability. This picture
would be indeed in agreement with the above-discussed RG results
of \cite{CL}. We consider further work in this direction highly
desirable, although it is clear that performing any perturbative
expansion around $N=\infty$ limit is expected to be a rather
technically challenging task.

We end up our presentation by giving a brief overview of a few
most recent advances in understanding the statistical mechanics of
a single particle in random high-dimensional potentials.
\subsection{Multiscale logarithmic potential}
As revealed by J.-P. Bouchaud and the present author in
\cite{FBprl,FBGREM}, the picture of potentials with short-ranged,
long-ranged, and logarithmic correlations presented above is still
incomplete, and misses a rich class of possible behavior that
survives in the thermodynamic limit $R \to \infty$. Namely, given
any increasing positive function $\Phi(y)$ for $0<y<1$, one can
consider potential correlation functions $f_V(u)$ in the
right-hand side of Eq.(\ref{1}) which take the following scaling
form
\begin{equation}\label{scalingln}
f_V(u)=-2 \ln{R}\,\,
\Phi\left(\frac{\ln{(u+a^2)}}{2\ln{R}}\right), \quad 0 \le u<R^2,
\end{equation}
This type of potential can be constructed by a superposition of
several logarithmically correlated potentials of the type
(\ref{2c}) with different cutoff scales $a_i$, and allowing those
cutoff scales to depend on the system size $R$ in a power-law way:
$a_i\sim R^{\nu_i},\, 0<\nu_i<1$ \cite{FBprl}.

 The thermodynamics of such system
in the limit $R\to \infty$ turns out to be precisely
equivalent\cite{FBprl} to that of the celebrated Derrida's
Generalized Random Energy Model (GREM)\cite{GREM,BoK}. The
REM-like case Eq.(\ref{2c}) turns out to
 be only a (rather marginal) representative of
this class: $\Phi(y)=g^2 y$.

The leading term in the equilibrium free energy turns out to be of
the form $F(T)=N \ln R \,{\cal F}(T)$, where for $0\le T\le
T_{AT}=\sqrt{\Phi'(1)}$
\begin{equation}
\label{freeenfin} -{\cal F}(T)=
T\nu_*(T)+\frac{\left[\Phi(\nu_*)-\Phi(0)\right]}{T}
+2\int_{\nu_*}^1\sqrt{\Phi'(y)}\, dy \,,
\end{equation}
where the parameter $\nu_*$ is related to the temperature $T$ via
the equation
\begin{equation}\label{main1}
T^2=\Phi'(\nu_*)\,.
\end{equation}
For $T>T_{AT}$ the free energy is instead given by
\begin{eqnarray}\label{freeensyma}
 -{\cal F}(T)=T+\frac{\left[\Phi(1)-\Phi(0)\right]}{T}\,.
\end{eqnarray}
 These expressions for the free energy can be given a clear
interpretation as describing a continuous sequence of "freezing
transitions" of REM type, with freezing happening on smaller and
smaller spatial scales\cite{FBprl,FBGREM}.

The form Eq.(\ref{freeenfin},\ref{freeensyma}) can give rise to a
rather rich multifractal behaviour of the Boltzmann-Gibbs
measure\cite{FBGREM}.  The associated singularity spectrum
$f(\alpha)$ calculated via Eq.(\ref{sinspec}) is positive in an
interval $\alpha\in(\alpha_{min},\alpha_{max})$, where the zeroes
$\alpha_{min},\alpha_{max}$ of the function $f(\alpha)$ are given
by
\begin{eqnarray}\label{minmax}\nonumber
\alpha_{min}=-\beta {\cal F}(\beta)-2\beta
\int_{0}^1\sqrt{\Phi'(y)}\,dy\,,\,\, \alpha_{max}=-\beta {\cal
F}(\beta)+2\beta \int_{0}^1\sqrt{\Phi'(y)}\,dy\,.
\end{eqnarray}
The singularity spectrum is symmetric with respect to the midpoint
of the interval of interest,
$\alpha_m=(\alpha_{min}+\alpha_{max})/2=-\beta {\cal F}(\beta)>0$,
where it has the maximum $f(\alpha_{m})=1$ as expected. Close to
this maximum, namely, in the subinterval
$\alpha\in(\alpha_{-},\alpha_{+})$ with $\alpha_{\pm}=\alpha_m\pm
2A(\beta)\frac{T}{T_{AT}}$, where $A(\beta)=\beta^2
(\Phi(1)-\Phi(0))$ the singularity spectrum has a simple parabolic
shape:
\begin{equation}\label{parab}
f(\alpha)=1-\frac{1}{4A(\beta)}\,(\alpha-\alpha_m)^2,\quad
\alpha_-\le \alpha\le \alpha_{+}\,.
\end{equation}
In particular, at the boundaries $f(\alpha_{\pm})=1-\beta_{AT}^2
\left(\Phi(1)-\Phi(0)\right)$. Note that in the REM-like limit
$\Phi(y)=g^2y$ we have $\alpha_{min/max}\to \alpha_{-/+}$ and the
parabolic behaviour is the only surviving, in agreement with the
results of \cite{2d,CL}.

At the same time outside the interval of parabolicity the general
GREM-like model shows a much richer multifractal structure
manifesting itself via a quite unusual behaviour of the
singularity spectrum close to the zeros
$\alpha_{min},\alpha_{max}$. To illustrate this fact, we consider
a broad class of functions $\Phi(y)$ behaving at small arguments
$y\ll 1$ as $\Phi(y)\approx C^2\,y^{2s+1}$ with $s\ge 0$ and the
coefficient $0<C<\infty$. In particular, in the limiting case
$s\to 0$ we are back to the old REM-like model. Now we can extract
the behaviour of the $f(\alpha)$ when approaching the endpoints
$\alpha_{min}$ or $ \alpha_{max}$. It is given by
\begin{eqnarray}
f(\alpha)\approx
\frac{s+1}{s^{s/(s+1)}}\,\alpha_{c}^{s/(s+1)}|\alpha-\alpha_{min/max}|^{\frac{1}{s+1}}\,,
\end{eqnarray}
where
\begin{eqnarray}
\alpha_c=\frac{2s^2}{(s+1)(2s+1)}(\beta
C\sqrt{2s+1})^{-\frac{1}{s}}\,.
\end{eqnarray}
We see that for any $s>0$ the derivative of the singularity spectrum
diverges as $f'(\alpha)\sim
|\alpha-\alpha_{min/max}|^{-\frac{s}{s+1}}\to \infty$. This is very
different from the standard behaviour observed in other disordered
systems \cite{ME,MG}: $f'(\alpha)<\infty$ at zeros of $f(\alpha)$.
At the level of multifractal exponents $\tau_q$ this feature is
translated to a rather unusual behaviour for large enough $|q|$,
namely: $\tau_q-q\alpha_{min}=-\alpha_{c}q^{-\frac{1}{s}}$ for
$q>T/T_{AT}$, and a similar formula for $q<-T/T_{AT}$. Note, that in
the standard situation one always observes linear behaviour
$\tau_q=q\alpha_{min,max}$ starting from some value of $|q|$, see
the formula (2.42) in \cite{ME} and discussions around it.

\subsection{Extrema of random landscapes and ergodicity breaking}

Another set of recent works on the random Gaussian model with
correlations specified by Eq.(\ref{1}) which deserves mentioning
is a continuing attempt\cite{FSW} to relate the phenomenon of
ergodicity breaking occuring at the level of statistical mechanics
to statistical properties of the minima (and other stationary
points) of high-dimensional Gaussian random surfaces ${\cal
H}({\bf x})$, see \cite{YFglass} for introduction to the
problematic. The authors managed to show that for a generic
smooth, concave confining potentials $V_{con}({\bf x})$ the
condition of the {\sf zero-temperature} replica symmetry breaking
coincides with one signalling that both mean total number of
stationary points in the energy landscape, and the mean number of
minima are exponential in $N$. For a generic system of this sort
the (annealed) complexity of minima vanishes cubically when
approaching the transition, whereas the cumulative annealed
complexity vanishes quadratically. One also can investigate how
the complexity depends on the index of stationary
points\cite{BrayDean,FSW}. In particular, in the vicinity of the
transition the saddle-points with a positive annealed complexity
must be close to minima, as they were found to have a vanishing
{\it fraction} of negative eigenvalues in the corresponding
Hessian.

\subsection{Statistical mechanics in a
sum of squared Gaussian-distributed potentials} Finally, let us
mention recent work \cite{mylast} on the statistical mechanics in
the energy landscape given by ${\cal H}({\bf x})=\frac{\mu}{2}{\bf
x}^2+\sum_{i=1}^K W^2_i({\bf x})$. Here $W_i({\bf x})$, with
$i=1,\ldots, K$ are assumed to be independent, identically
distributed Gaussian functions with zero mean, the variance
$\langle W_i^2\rangle = \sigma\left(\frac{{\bf x}^2}{N}\right)$
and the structure function $\left\langle \left[W_i\left({\bf
x}_1\right)- W_i\left({\bf
x}_2\right)\right]^2\right\rangle=2\phi_W\left(\frac{({\bf
x}_1-{\bf x}_2)^2}{N}\right)$. Note important differences from the
Gaussian case: (i) the absence of the factor $N$ in front of the
(co)variance, in contrast to Eq.(\ref{1}), and (ii) necessity of
specifying both functions $\sigma$ and $\phi_W$, as the phase
diagram will actually depend on both of them, in contrast to the
the discussion in the footnote $1$. The free energy of such a
system turns out to have a well-defined large$-N$ limit provided
we scale $K=N\kappa$, and consider the parameter
$0<\kappa<\infty$. Naively one may think that the central limit
theorem (CLT) would imply that the sum of $K=O(N)$ random terms
effectively behaves as a Gaussian potential. A thorough
consideration shows that such a reasoning is however deficient for
the statistical mechanics problem in hand. Indeed, with lowering
the temperature deep minima of the resulting potential start
playing most prominent role, and the description of those minima
goes beyond the applicability of CLT. This fact suggests that the
statistical mechanics of such model may have features rather
different from the former Gaussian case due to different
statistics of deep minima\cite{BM}. The dynamics in this case may
also be rather different, see interesting related results in
\cite{Dean}.

The free energy can be evaluated in the limit $N\to \infty$ by
extending the methods of \cite{FS}, and the system shows both
similarities and dissimilarities to the Gaussian case. In
particular, the difference between the short-range and long-range
potentials remains to be important, but manifests itself in a
somewhat different way. One again finds that the replica-symmetric
solution is unstable at low enough temperatures. Let us discuss
here only the simplest case of a short-ranged potential with
position-independent variance $\sigma\left(\frac{{\bf
x}^2}{N}\right)=\sigma\equiv f_W(0)$, where $f_W(u)$ stands for
the covariance function of the field $W$, related to the structure
function as $\phi_W(u)=f_W(0)-f_W(u)$. The equation for the
transition (de-Almeida-Thouless) line $T_{AT}(\mu)$ is then given
in terms of the structure function by:
\begin{equation}\label{ATG2}
\frac{\left[\phi_W'(\tau_{AT})\right]^2-\phi_W''(\tau_{AT})\left[\sigma-\phi_W(\tau_{AT})\right]}
{\left[1+\frac{\phi_W(\tau_{AT})}{T_{AT}}\right]^2}=\frac{\mu^2}{\kappa},
\quad \tau_{AT}=T_{AT}/\mu
\end{equation}
By investigating this expression one finds that the phase with
broken replica symmetry may exist only as long as the parameter
$\kappa$ exceeds some critical value
\begin{equation}\label{ATcond}
\kappa>\kappa_{cr}=\frac{1}{1+\frac{f_W(0)f_W''(0)}{f_W'^{2}(0)}}\,.
\end{equation}
Moreover, for every such $\kappa$  the curvature of the confining
potential must satisfy the inequality
\begin{equation}\label{ATcond1}
\mu<\mu_{cr}=\sqrt{\kappa[f_W(0)f_W''(0)+f_W'^{2}(0)]}+f_W'(0)\,.
\end{equation}

In the case of long-range potentials one has to take into account
the fact of position-dependent variance, which makes the analysis
more complicated, and the corresponding phase diagram quite
intricate. These features are currently under
investigation\cite{mylast}.\\

{\bf Acknowledgements}. This research was supported by Bessel
award from Humboldt foundation, and by grant EP/C515056/1 from
EPSRC (UK). The author is grateful to J.P. Bouchaud,  H.-J.
Sommers, and I. Williams for their collaboration on various
aspects of the problems discussed in this presentation. The author
also appreciates kind hospitality extended to him during his
prolonged visits to the Institute of Theoretical Physics, Cologne
University, Germany, where the major part of the reported results
had been obtained.


\begin{thebibliography}{99}

\bibitem{ME} F. Evers and A.D. Mirlin,  e-preprint arXiv:0707.4378 [cond-mat.mes-hall]

\bibitem{MG} C. Monthus and T. Garel, {\it Phys. Rev. E} {\bf 75}
(2007), Art. No. 051122

\bibitem{2d} H. E. Castillo, C.C. Chamon, E. Fradkin, P.M. Goldbart
and C. Mudry, {\it Phys. Rev. B } {\bf 56}, 10668 (1997)


\bibitem{MP} M.\, Mezard and G.\, Parisi {\it J.Phys.A:Math.Gen.} {\bf 23} (1990), L1229;
 {\it J.Phys.I France} {\bf 1}, (1991)
809; {\it J.Phys.I France} {\bf 2}, (1992) 2231;

\bibitem{Engel} A.\, Engel {\it Nucl.Phys.B} {\bf 410} (1993),617


\bibitem{FM} S.\, Franz and M.\, Mezard {\it Physica A}, {\bf 210}
(1994), 48

\bibitem{longtime} L. F. Cugliandolo and P. Le Doussal {\it
Phys.Rev.E } {\bf 53}, (1996) 1525

\bibitem{BM}  J.P. Bouchaud and M. Mezard, {\it J. Phys. A: Math.
Gen}{\bf 30} (1997), 7997

\bibitem{FS} Y. V. Fyodorov and H.-J. Sommers {\it Nucl.Phys.B [FS]}
{\bf 764}, 128 (2007)


\bibitem{CDT}  D. M. Carlucci, C. de Dominicis, and T. Temesvari
 {\it J.Phys. I France}, {\bf 6} (1996), 1031

\bibitem{AT} J.R.L. de Almeida and D.J.Thouless, {\it J.Phys.A}
{\bf 11} (1978), 983

\bibitem{CL} D. Carpentier, P. Le Doussal
{\it Phys.Rev.E} {\bf 63}, 026110 (2001)

\bibitem{DS} B. Derrida and H. Spohn {\it J. Stat. Phys.} {\bf 51} 817, (1988)

\bibitem{Derrida} B. Derrida {\it Phys.Rev.B} {\bf 24}, 2613 (1981)

\bibitem{Moore} M. Moore {\it Phys. Rev. Lett.} {\bf 96}, 137202
(2006)

\bibitem{FBprl} Y. V. Fyodorov and J. P. Bouchaud , {\it JETP Letters}, {\bf 86}, 487 (2007)
[e-preprint arXiv:0706.3776 ].

\bibitem{FBGREM} Y. V. Fyodorov and J. P. Bouchaud, e-preprint
arXiv:0711.4006 [cond-mat.dis-nn]

\bibitem{GREM} B. Derrida {\it J. Phys. Lett.} {\bf 46}, 401 (1985); B.
Derrida and E. Gardner {\it J. Phys. C} {\bf 19} 2253 (1986) and
{\bf 19} 5783 (1986)

\bibitem{BoK} A. Bovier and I. Kurkova, {\it Ann. I.H. Poincare} -
PR {\bf 40}, 439 (2004), and PR {\bf 40}, 481 (2004)

\bibitem{FSW} Y. V. Fyodorov and I. Williams, {\it J. Stat. Phys.} {\bf 129}, 1081 (2007) [arXiv:cond-mat/0702601]
; Y. V. Fyodorov, H.-J. Sommers, and I. Williams, {\it JETP
Letters} {\bf 85}, 261 (2007)

\bibitem{YFglass} Y. V. Fyodorov  {\it Phys. Rev. Lett.} {\bf 92}, Art. No. 240601 (2004); Erratum: {\it ibid.}
{\bf 93}, Art. No. 149901 (2004) and  Acta Physica Polonica B,
{\bf 36}, 2699 (2005)

\bibitem{BrayDean} A. J. Bray and D. S. Dean, {\it Phys. Rev. Lett.} {\bf 98}, Art. No. 150201 (2007)

\bibitem{mylast} Y. V. Fyodorov, in progress.

\bibitem{Dean} C. Touya and D. S. Dean {\it J.Phys.A:Math.Gen.} {\bf 40} (2007),
919

\end{thebibliography}
\end{document}